\documentclass[12pt]{article}
\usepackage{axodraw}
\usepackage{epsfig}   
\usepackage{amssymb}
\usepackage{latexsym}
\usepackage{amsmath}

\usepackage{graphics,subfigure} 

\usepackage{color} 

\newcommand{\lsim}{\raisebox{-0.13cm}{~\shortstack{$<$ \\[-0.07cm] $\sim$}}~}
\newcommand{\gsim}{\raisebox{-0.13cm}{~\shortstack{$>$ \\[-0.07cm] $\sim$}}~}
\newcommand{\bea}{\begin{eqnarray}}
\newcommand{\eea}{\end{eqnarray}}
\newcommand{\beq} {\begin{equation}}
\newcommand{\eeq} {\end{equation}}

\begin{document}
\pagestyle{empty}
\begin{flushright}
\today
\end{flushright}
\begin{center}
{\large\sc {\bf Radius of the innermost stable circular orbit of black-holes 
with localized Astrophysical Dark Matter halo
}}
\end{center}
\vspace{1.0truecm}
\begin{center}
{\large  Anup S. Karekar~\footnote{Email-id:~karekaranup28@gmail.com} and 
	Siba Prasad Das~\footnote{Email-id:~spd.phy@unishivaji.ac.in}}\\
\vspace*{5mm}
{}$^1${\it  
Department of Physics, Rajaram College \\[0.15cm]
Vidyanagar, Kolhapur-416004, Maharashtra, India} \\[0.07cm]
{}$^2${\it 
Department of Physics, Shivaji University Kolhapur,\\[0.15cm]
Vidyanagar, Kolhapur-416004, Maharashtra, India} \\[1.0cm]
\end{center}

\begin{abstract}
        We investigate the geodesic structure of a static, spherically symmetric black-hole in the 
	background of a localized Astrophysical Dark Matter (ADM)-halo. The halo is modeled by a 
	core-radius ($r_0$), regularized density profile that vanishes at the event horizon 
	and exponential factor ensures the finite total halo mass. We solved the Einstein field 
	equations numerically with equation of state with $p=w\rho$ for stiff matter $w=1$ scenario. 
	By analyzing the effective potential and its first and second derivatives, we evaluated the 
	characteristic radius of the time-like innermost stable circular orbit ($r_{\text{ISCO}}$) 
	of massive particle in the vicinity of hairy black-hole. Our numerical analysis is consistent 
	with the recently proposed theoretical upper bound $r_{\text{ISCO}} \le 6$ normalized to 
	unit mass black-hole. We showed that the ADM-halo shifts $r_{\text{ISCO}}$ inward with respect 
	to the Schwarzschild (bald) scenario and the shifts are larger for larger halo core-radius. 
        In particular our numerical finding enumerate $r_{\text{ISCO}}$ are approximately  
	$5.97$, $5.61$, $5.53$ for $r_0=2.0,10.0,20.0$ respectively. We believe that our analysis 
	with ADM-halo model shed light on probing black-hole hair through high-energy 
	astrophysical observables.
\end{abstract}

\newpage
\setcounter{page}{1}
\pagestyle{plain}

\section{Introduction}
\label{sec:intro}

The first image of $M87^{*}$ by the Event Horizon Telescope~\cite{EventHorizonTelescope:2019dse}, have 
strengthened the observational evidence of general relativity and motiving the deeper understanding 
of the geometry of the black-hole. The most important geometry related to the black-hole physics are the 
photon sphere and the innermost stable circular orbit (ISCO). The photon sphere determines the black-hole shadow 
and satisfy several universal bounds~\cite{Hod:2013jhd,Hod:2018mvu,Hod:2020pim,Peng:2020shc,Hod:2023jmx,
Liu:2024odv}.The ISCO for the massive(time-like) particles plays a central role in accretion disk physics and 
gravitational-wave astronomy. The ISCO radius ($r_{ISCO}$) is well understood for Schwarzschild and Kerr 
black-holes, for example studied in ~\cite{Teo:2020sey}. However more general solutions containing 
external matter fields, such as Reissner--Nordstrom black-hole~\cite{Wald:1984rg} and compact fluid 
configurations~\cite{Tolman:1939jz,Buchdahl:1959zz,Ivanov:2002jy} exhibit richer near-horizon 
structures that can significantly modify the particle dynamics. In ~\cite{Hod:2013jhd} the 
universality of photon-sphere has been obtained to be $r_{\gamma} \lsim 3M$. In \cite{Cen:2025atm} 
the upper limit of the $r_{ISCO}$ for massive-particle orbits in generic black-hole space-times have 
been derived analytically and put a stronger theoretical upper limit, $r_{ISCO} \lsim 6M$.

The Astrophysical Dark Matter(ADM)-halo density profile is parameterized by a central dark-matter density ($\rho_0$) and 
the halo core radius($r_0$). There is as such no single universal observational value 
of $\rho_0$ rather the values has been fitted from the galaxy, halo mass, and the specific 
density profile used. Importantly the recent rotation-curve\cite{rotation} and weak-lensing studies\cite{weak} from 
dwarf, spirals and ellipticals galaxies shows a remarkable scaling relation\cite{donato}:

\begin{equation}
\rho_0 r_0 \simeq
10^{2.15 \pm 0.2}
M_\odot {\rm pc}^{-2},
\end{equation}

In our analysis we studied how the location of $r_{ISCO}$ for the massive-particle modifies in 
presence of the ADM-halo, mimicking the presence of static hairy 
black-hole~\cite{Herdeiro:2014goa,Herdeiro:2015waa,Hui:2019aqm} or 
Dark-matter density spike model\cite{Chan:2022gqd,Chan:other}.

The plan of this paper is as follows. In the next section we will describe briefly  
the Astrophysical Dark Matter (ADM)-halo profile. In Sec.3 we discuss the 
Hairy black-hole with ADM-halo model. In Sec.4 we analyze the criterion of obtaining 
$r_{ISCO}$ from the effective potential for the massive(time-like) particle. Our 
numerical finding is discussed in Sec.5. In Sec.6 we summarize our analysis and mention 
possible future outlook.

\section{Astrophysical Dark Matter (ADM)-halo}
\label{sec:dmhalo}

In our analysis we have considered the ADM-halo energy density profile as:

\begin{equation}
	\label{density}
	\rho(r)=\rho_0 \left(1-\frac{r_H}{r}\right) \exp \left[-\frac{(r-r_H)}{r_0}\right] ~if~r \geq r_H~ else~\rho(r)= 0,
\end{equation}

where $\rho_0$ is the characteristic ADM-halo energy density, $r_H$ is the black-hole event horizon radius, 
and $r_0$ is the characteristic halo core radius. We will see in our numerical analysis that this density 
profile are very much important for getting the exact location of $r_{ISCO}$. This particular ADM-halo profile vanishes 
exactly at the event horizon, i.e., $\rho(r_H)=0$ and avoids the appearance of divergent 
densities near the horizon and provides a mathematically regular description of the ADM-halo  
density profile around the black-hole. The density dropped rapidly due to the presence of the 
the exponential factor, for $r \gg r_0$; hence the ADM-halo cloud remains localized 
around the black-hole and possesses a finite total ADM-halo mass. Our analysis shed light on 
how a localized external matter distribution modifies the location of strong-field circular geodesics 
around a non-rotating hairy black-hole.

The most important quantity influencing the particle dynamics is the enclosed ADM-halo mass:

\begin{equation}
	M_{ADM}(r)=4\pi
\int_{r_H}^{r}
\rho(r'),r'^2,dr'.
\end{equation}

We evaluated the integral numerically and then deduce the $\mu$, $\delta$-parameters appear in 
the metric functions and ISCO equations. For the interpretations of black-hole hair, the 
density naturally represents a compact dark-matter cloud concentrated outside the horizon. Therefore, 
this can be interpreted as an effective form of black-hole hair rather than a large-scale galactic halo.

Although mathematically convenient, the factor, $(1-\frac{r_H}{r})$ is not directly motivated 
by standard dark-matter simulations. In many dark-matter spike models, $\rho(r)\propto r^{-\gamma}$,
with $1\lesssim \gamma \lesssim 2.5$, implies that the density increases toward the central region.
Therefore, the proposed profile in our analysis differs from conventional dark-matter spike scenarios.
A possible interpretation of the presence of $(1-\frac{r_H}{r})$ is the depletion of dark-matter 
particles due to accretion and capture by the black-hole. This is important to mention 
here that in the limit of either $\rho_0 \to 0$ or $r_0 \to 0$, the model leads to the 
classical Schwarzschild (bald) black-hole scenario.

For galaxies~\cite{galaxy}, the central dark-matter density is typically $10^{-3}\lesssim \rho_0 \lesssim 1
\quad M_\odot\,\mathrm{pc}^{-3}$. For Dwarf spheroidal galaxies the corresponding range is 
$0.1\lesssim \rho_0 \lesssim 1 \quad M_\odot\,\mathrm{pc}^{-3}$.  The conversion 
factor is $1\,M_\odot\,\mathrm{pc}^{-3} = 6.77 \times 10^{-20}\,\mathrm{kg\,m^{-3}}$. For ISCO calculations around 
a hairy black-hole, a reasonable parameter scan would be 
$\rho_0= \left\{ 10^{-5} -to- 1 \right\} M_\odot\,\mathrm{pc}^{-3}$. 
In our analysis we used the natural unit with $G_N = c =1$ with unit normalized black-hole mass, i.e.,  
$M_{BH}$=1. We took $\rho_0= 3.3144 \times 10^{-5}$ in natural unit in our analysis which is in 
the ball-park values for ADM-halo model.

\section{The model: Hairy black-hole with ADM-halo}
\label{sec:hbhadm}

We have assumed the static, spherically symmetric, and asymptotically flat 
black-hole space-time with ADM-halo density is in the background. The importance of 
black-hole hair~\cite{Herdeiro:2014goa,Herdeiro:2015waa,Hui:2019aqm} and dark-matter 
density spike model have been studied in detail~\cite{Chan:2022gqd,Chan:other}.

In our analysis the line-element of the static, spherically symmetric, and asymptotically flat black-hole 
space-time metric can be written as ~\cite{Hod:2013jhd,Hod:2020pim,Nunez:1996xv}
\begin{align}
        \label{duguinew}
        ds^2=-e^{-2\delta}\mu dt^2+\mu^{-1}dr^2+r^2(d\theta^2+\sin^2\theta d\phi^2)\;,
\end{align}
where $\mu(r)$ and $\delta(r)$ are depend upon the radial co-ordinate $r$. The values of 
$\delta(r)$ is non-zero in our analysis thus it is applicable to hairy 
black-hole~\cite{Volkov:1998cc,Volkov:2016ehx}. Our space-time metric is asymptotically flat: 

\begin{align}
\label{as}
        \mu(r\rightarrow \infty)\rightarrow 1\quad\mathrm{and}\quad\delta(r\rightarrow \infty)\rightarrow 0\;.
\end{align}

We take the component of the stress-energy tensor has the following diagonal form 
$T= T^t_t, T^r_r,T^{\theta}_{\theta},T^{\phi}_{\phi} = (-\rho,p,p_T,p_T)$ in the $t,r,\theta,\phi$ 
spherical polar co-ordinates. The parameter $\rho$, $p$ and $p_T$ are the energy density,  
radial pressure, and tangential pressure respectively.

The mass $m(r)$ enclosed within a sphere of radius $r$ is expressed as
\begin{align}
\label{m}
	m(r)=\frac{1}{2}r_H+ M_{ADM}(r)\;.
\end{align}
and the horizon mass $m(r_H)$ is equal to $\frac{r_H}{2}$. Considering the conditions of finite mass 
configuration we have checked that:  
\begin{align}
\label{rho}
        r^3\rho(r)\rightarrow0\ \mathrm{when}\  r\rightarrow \infty \;.
\end{align}

From Einstein field equations $G^{\mu}_{\nu}=8\pi T^{\mu}_{\nu}$, one can find~\cite{Cen:2025atm}:
\begin{align}
\label{mup}
&\mu'=-8\pi r \rho+\frac{1-\mu}{r}\;,\\
\label{deltap}
&\delta'=-\frac{4\pi r(\rho+p)}{\mu}\; = -\frac{4\pi r\rho(1+w)}{\mu}\;= -\frac{8\pi r\rho}{\mu}\;,
\end{align}
where the prime indicates the derivative with respect to the radial coordinates $r$ and 
we consider the equation of state $p= w\rho=\rho$, i.e., $w=1$ for the stiff matter distribution.

The $\mu$ parameter in the metric has the following form:

\begin{align}
\label{murm}
\mu(r)=1-\frac{2m(r)}{r}\;,
\end{align}

and has been used while doing the numerical integration of Eqns.\ref{mup},\ref{deltap}.

We have been checked that the second derivatives of $\mu(r)$ and $\delta(r)$ 
with respect to $r$ have the following form:
\begin{align}
	\label{mu2}
	&\mu''=\frac{-1+\mu-r[\mu'+8\pi r(\rho+r\rho')]}{r^2}\;,\\
	\label{delta2}
	&\delta''=-\frac{4\pi [-r\rho \mu'+p(\mu-r\mu')+\mu(\rho+rp'+r\rho')]}{\mu^2}\;.
\end{align}

We have been solved the Eqns.\ref{mu} and \ref{delta} using numerical integration to get 
the correct analytical expression of $\mu(r)$ and $\delta(r)$. The integration constants 
have been fixed with proper boundary conditions of asymptotic flatness of space-times 
following Eqn.\ref{as}. Moreover we demand the regularity at the event horizon($r = r_H$) and 
this enforces: 

\begin{align}
        \label{miunew}
        \mu(r_H)=0\quad\mathrm{with}\quad\mu'(r_H)\geq 0\;,
\end{align}
and
\begin{align}
        \label{detanew}
        \delta(r_H)<\infty\quad ,\quad\delta'(r_H)<\infty \;.
\end{align}

\section{Innermost stable circular orbit(ISCO)}
\label{sec:isco}

Due to the spherical symmetry we have consider the particle is moving in 
the equatorial plane, i.e.,, $\theta=\frac{\pi}{2}$. The Lagrangian describing the 
geodesics in Eqn.\ref{duguinew} is~\cite{Cen:2025atm}: 
\begin{align}
	\label{lagran}
	2\mathcal{L}=-e^{-2\delta}\mu \dot{t}^2+\mu^{-1}\dot{r}^2+r^2\dot{\phi}^2=\epsilon \;,
\end{align}
where $\epsilon=-1,0,+1$ for time-like(massive), photon-like(massless) and space-like geodesics, 
respectively. To analyze ISCO for massive particle we take $\epsilon=-1$. 
As the metric in Eq.\ref{duguinew} is independent of both $t$ and $\phi$ so there are two 
conserved quantities along the geodesic, i.e. energy $E$ and angular momentum $L$.

The generalized momenta has been derived from the Lagrangian in Eqn.\ref{lagran} as follows:  
\begin{align}
	\label{ppt}
	&p_t=-e^{-2\delta}\mu\dot{t}=-E\;,\\
	\label{pphi}
	&p_{\phi}=r^2\dot{\phi}=L\;,\\
	&p_r=\mu^{-1}\dot{r}\;.
\end{align}

After substituting Eqs.\ref{ppt} and \ref{pphi} into Eq.\ref{lagran}, 
one can finds
\begin{align}
	\label{dotr}
	\dot{r}^2=\frac{E^2}{e^{-2\delta}} - \mu(\frac{L^2}{r^2}+1)
\end{align}
the geodesics for the massive particle.

The effective potential of the geodesic and its first derivative for the massive particle can be defined as
\begin{align}
	\label{veff}
	V_{\mathrm{eff}}(r)=\frac{E^2}{e^{-2\delta}} - \mu(\frac{L^2}{r^2}+1)\;,
\end{align}

\begin{align}
	\label{vpeff}
	V'_{\mathrm{eff}}(r)=\frac{\mu'}{\mu}(V_{\mathrm{eff}} - E^2 e^{2\delta}) +2(E^2 e^{2\delta}\delta'+\frac{L^2\mu}{r^3}) \;.
\end{align}

As it is well understood that the circular orbits satisfy two conditions i.e., 
$V_{\mathrm{eff}}(r)=0$ and ${V}_{\mathrm{eff}}'(r)=0$. Therefore from the Eqns.\ref{veff},\ref{vpeff} we 
can find the solutions of the energy ($E$) and the angular momentum ($L$) of the massive particle's circular 
orbit: 
\begin{align}
	\label{E2exp}
	E^2=\frac{e^{-2\delta}\mu^2}{\mu+r\mu \delta'-r\mu'/2}\;,
\end{align}
and
\begin{align}
	\label{l2exp}
	L^2=\frac{\mu r^2}{\mu(1+r\delta') -r\mu'/2} -r^2\;.
\end{align}

These values of $E^2$ and $L^2$ from Eqns.\ref{E2exp},\ref{l2exp} have been passed in the expression 
of $V''_{\mathrm{eff}}$ in Eqn.\ref{vppeff} to get $r_{ISCO}$.

To find the ISCO, is also well understood, one put the additional condition, i.e., the second 
derivative of $V_{\mathrm{eff}}(r)$ should be zero, $V_{\mathrm{eff}}''(r)=0$. 
We find, similar to ~\cite{Cen:2025atm}, the second derivative of $V_{\mathrm{eff}}(r)$ has the following form:
\begin{align}
	\label{vppeff}
	V''_{\mathrm{eff}}=\frac{4L^2\mu'+2e^{2\delta}E^2r^3(2\delta'^2+\delta'')-r(L^2+r^2)\mu''}{r^3} -\frac{6 L^2\mu}{r^4}\;.
\end{align}

Now it is very clear that the position of the $r_{ISCO}$, shown in the Sec.\ref{sec:numerical}, is where  $V_{\mathrm{eff}}''(r)=0$. 

\section{Numerical Analysis}
\label{sec:numerical}

\begin{figure}[ht!]
\begin{center}
\raisebox{0.0cm}{\hbox{\includegraphics[angle=0,scale=0.41]{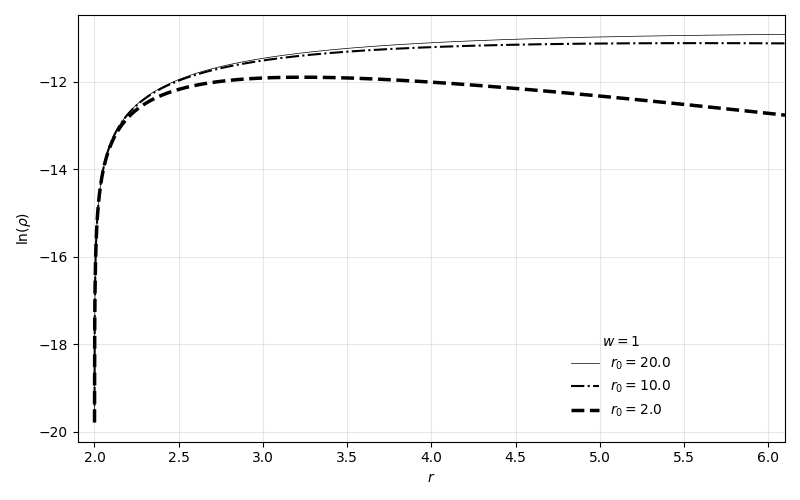}}}
\raisebox{0.0cm}{\hbox{\includegraphics[angle=0,scale=0.41]{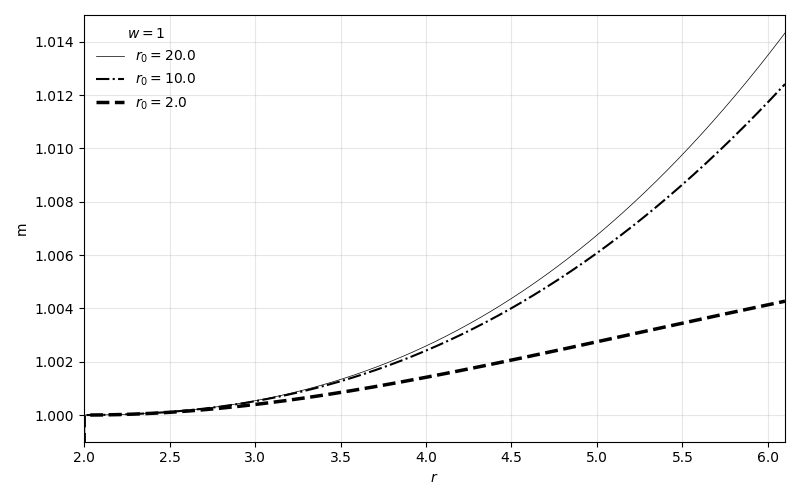}}}
	\caption{In the left-panel the ADM-halo $ln(\rho(r))$ is shown 
	as a function of radial co-ordinates. In the right-panel we plotted the mass 
	profile enclosed within the radial distance, $m(r)$, of the black-hole 
	together with the ADM-halo for three different values of halo-core radius 
	$r_0=2.0,10.0$ and $20.0$ using dashed, dotted-dashed and solid-plot   
	style respectively. In the right-panel it seems the halo mass is increasing, however 
	we checked that the halo mass has been saturated for $r \simeq 20(150)$ with core-radius 2.0(20.0).}
\label{rhomass}
\end{center}
\end{figure}

We have shown the density profile, i.e., Eqn.\ref{density} of the ADM-halo as a function 
of the radial co-ordinate in the left-panel of Fig.\ref{rhomass}. The presence of numerical 
factor $(1 -\frac{r_H}{r})$ ensures the density becomes zero at the event horizon 
and its exponentially decaying. Since we have been used log-profile it shows the 
negative values and with larger(smaller) core-radius the density profiles 
shifts in the smaller(larger) negative values.

In the right-panel of Fig.\ref{rhomass} we have shown the masses of the black-hole with ADM-halo. 
Clearly for $r=r_H$ the masses of the ADM-halo is zero and black-hole has unit masses following 
$r = 2m$ the Schwarzschild  condition. With increasing values of the radial distances 
the ADM-halo has acquiring more masses so the total mass is increasing. For larger 
core-radius the masses are diffused more so to reach the plateau it needs larger 
distances. While for smaller core-radius the masses are more concentrated close 
to the event horizon. With $r_0=2.0$ the right-panel shows that the masses increases 
linearly with radial co-ordinate.

\begin{figure}[ht!]
\begin{center}
\raisebox{0.0cm}{\hbox{\includegraphics[angle=0,scale=0.51]{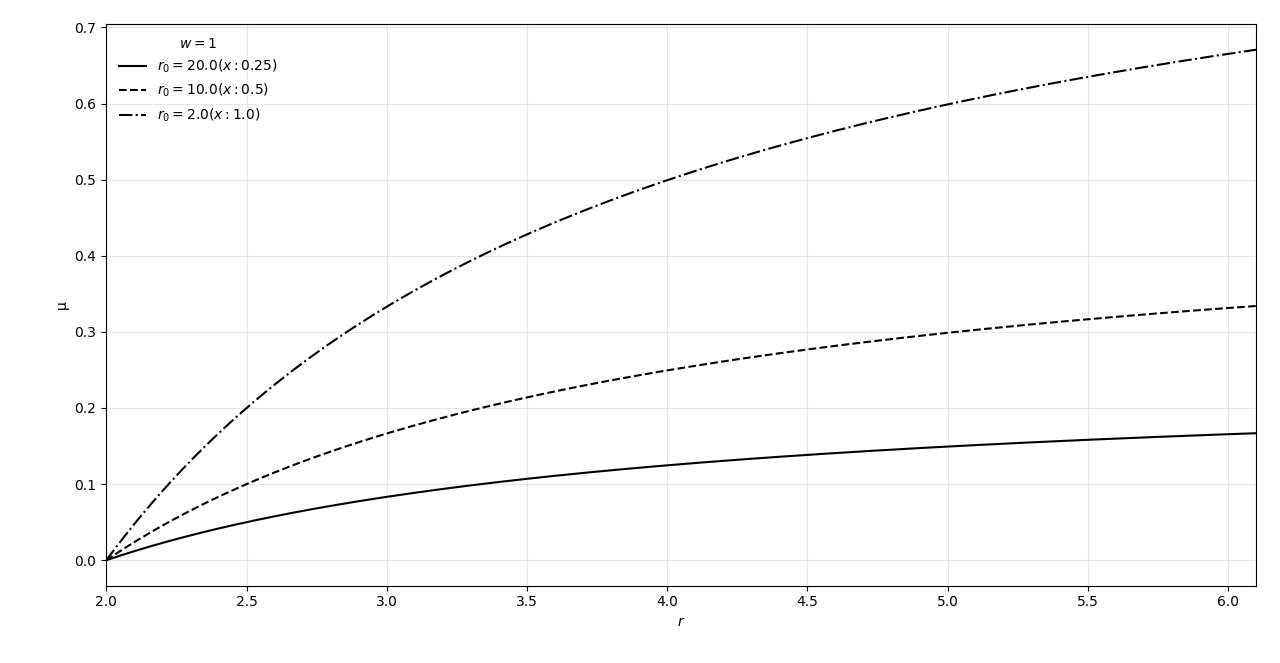}}}
\raisebox{0.0cm}{\hbox{\includegraphics[angle=0,scale=0.41]{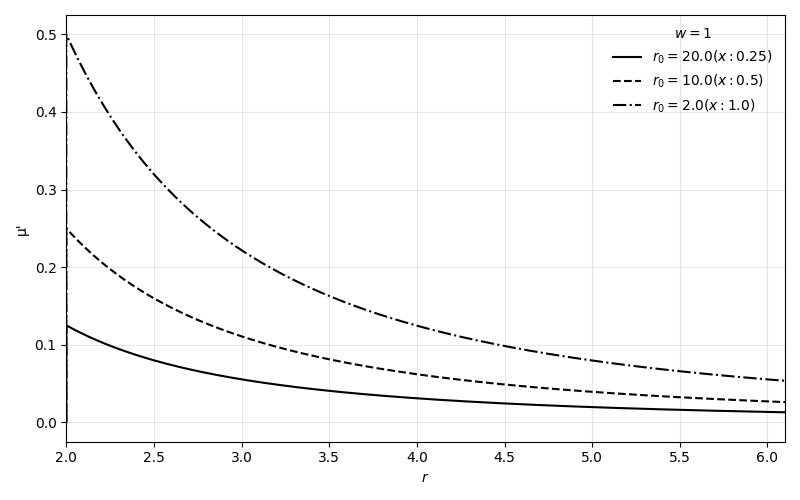}}}
\raisebox{0.0cm}{\hbox{\includegraphics[angle=0,scale=0.41]{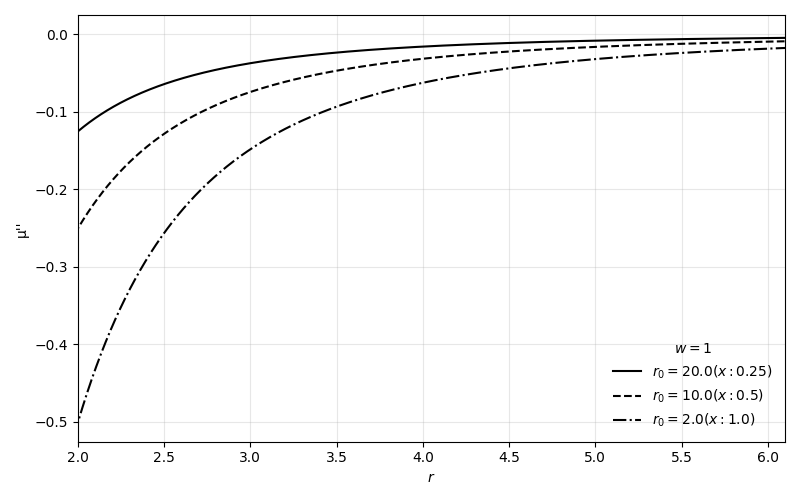}}}
	\caption{The $\mu$ (top), the first derivative $\mu'$ (bottom-left) 
	and the second derivative $\mu''$ (bottom-right) 
	as a function of the radial coordinates ($r$) for black-hole 
	with ADM-halo for three different core radius $r_0$=2.0(x:1.0),10.0(x:0.50) and 20.0(x:0.25)  
	with different scale-factors mentioned in the parenthesis using dashed, 
	dotted-dashed and solid-plot style respectively.}
\label{mu}
\end{center}
\end{figure}

In Fig.\ref{mu} we have shown the $\mu(r)$ (top-panel), $\mu'(r)$ (bottom left-panel) 
and $\mu''(r)$ (bottom right-panel) as a function of radial coordinates for three 
different choices of $r_0=20.0$(x=0.25), $10.0$(x=0.50), $2.0$(x=1.0) with different 
scale-factors(x) mentioned in the corresponding parenthesis. 
The $\mu(r)$ profile, given the black-hole with the ADM-halo contribution inside that radius, show that it is 
attractive in nature as expected. For larger values of $r_0=20.0$ ($10.0$) the $\mu(r)$  
is more flatter than the lower values of $r_0=10$($2.0$) as lower values the mass is more 
concentrated near the event horizon disk thus the $\mu(r)$ is much more 
steeper in the lower values of $r_0$.

It seems $\mu' > 0$ (in the bottom left-panel) as we go outward, the space-time becomes less 
strongly curved (as gravity weakens with distances). It is clear that lower values of 
the $r_0$ the $\mu'$ is much larger than the higher values of $r_0$. In the vicinity 
of the horizon the $\mu'$ (it's a kind of effective attractive forces) are much stronger 
and for $r_0=2$ is the values of $\mu'$ are much larger than the other two values of $r_0=10$ and $20$.
The bottom-right panel shows the profile of $\mu''$ and the bending of the profile 
happens at a decreasing rate. The $\mu''$ is very strong near the black-hole while in far away it is nearly flat. 

All these three parameters of Fig.\ref{mu} and their properties discussed above are inherited 
from the mass profile $m(r)$ and the positivity of the energy density. 

\begin{figure}[ht!]
\begin{center}
\raisebox{0.0cm}{\hbox{\includegraphics[angle=0,scale=0.27]{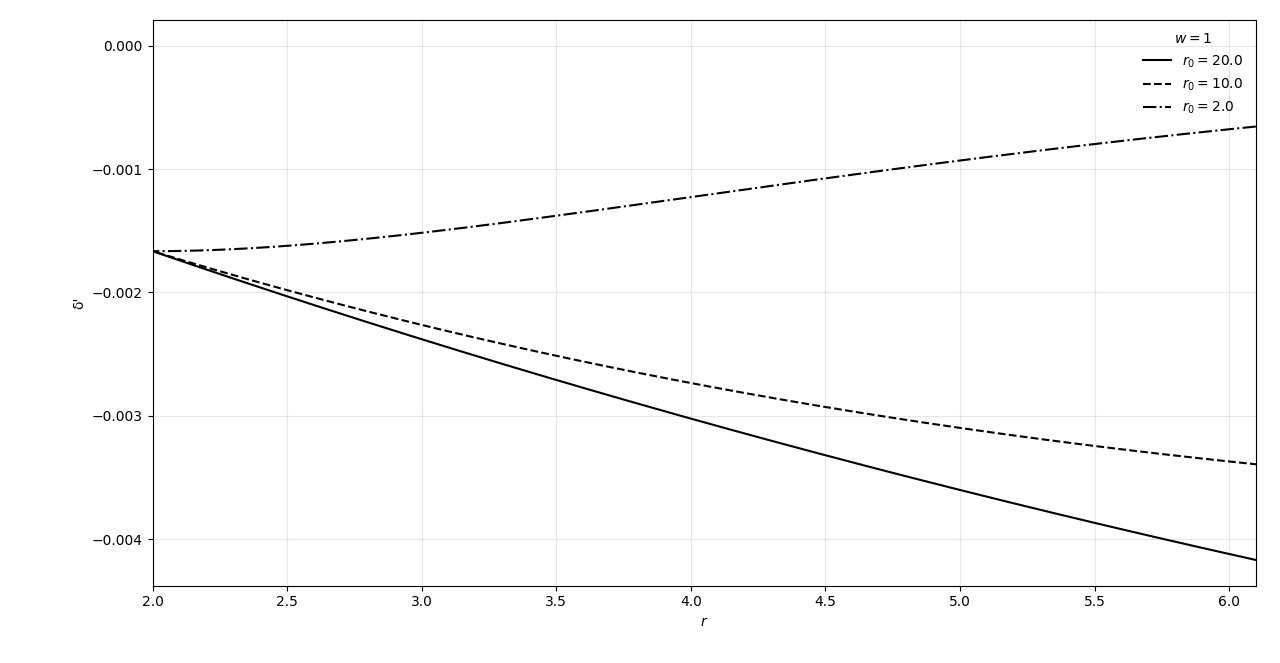}}}
\raisebox{0.0cm}{\hbox{\includegraphics[angle=0,scale=0.39]{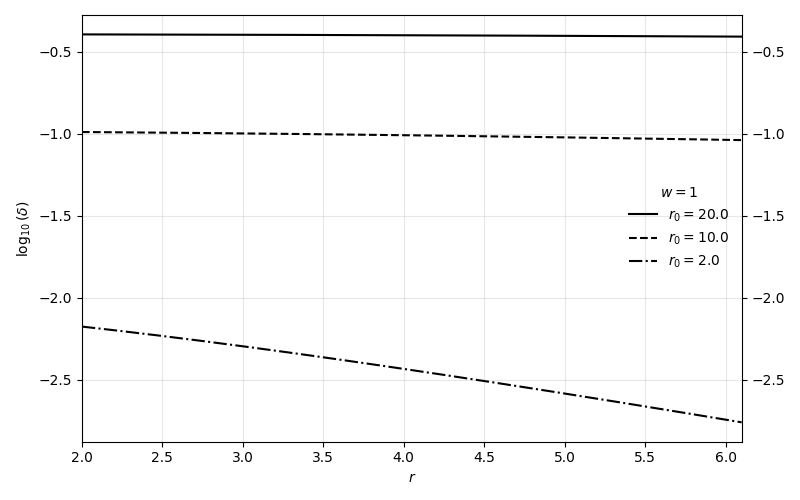}}}
\raisebox{0.0cm}{\hbox{\includegraphics[angle=0,scale=0.41]{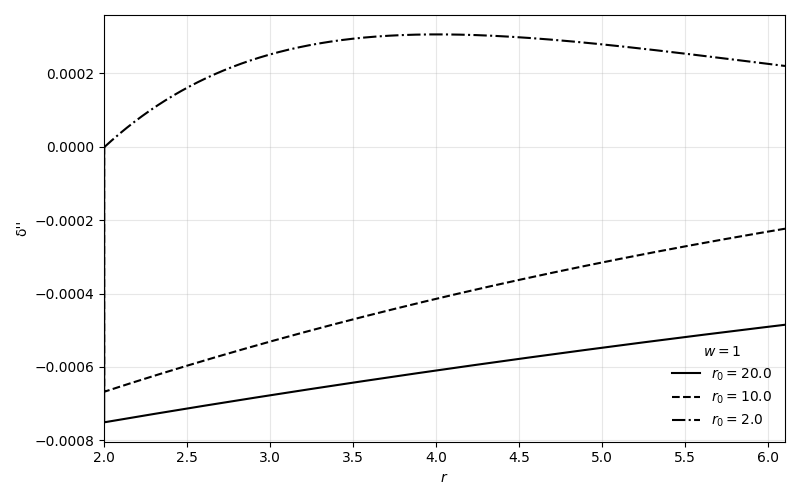 }}}
\raisebox{0.0cm}{\hbox{\includegraphics[angle=0,scale=0.41]{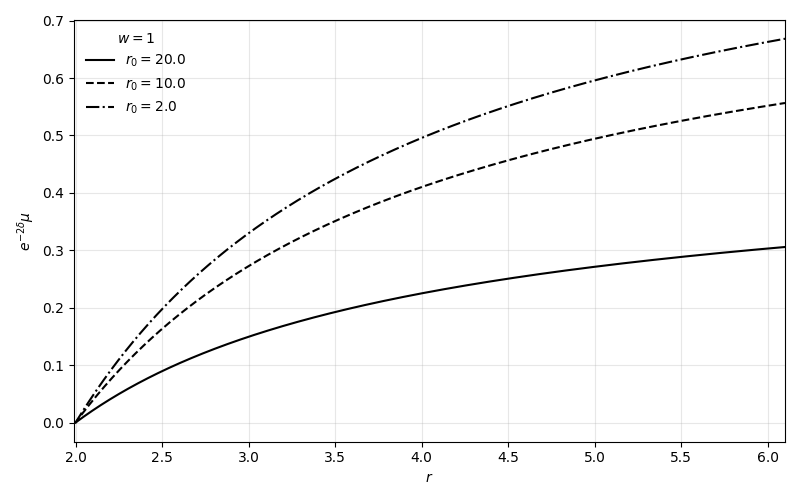}}}
	\caption{ The $\delta'$ (top-left), $log_{10}\delta$(top-right), $\delta''$
	(bottom-left) and  $e^{-2\delta}\mu$ (bottom-right)   
	profiles as a function of the radial coordinates($r$) for the black-hole with 
	ADM-halo for three different values of core-radius $r_0=2.0$, $10.0$ and $20.0$ 
	using dashed, dotted-dashed and solid-line style respectively.}
\label{delta}
\end{center}
\end{figure}

We have shown the variations of $\delta'(r)$, $log_{10}\delta(r)$, $\delta''(r)$ and $e^{-2\delta(r)} \mu(r)$  
as a function of radial coordinates in Fig.\ref{delta}. The parameter $\delta$ is effectively 
the red-shift parameter of the hairy black-hole.

In the top-left panel of Fig.\ref{delta} we have shown the variations of $\delta'(r)$ as a 
function of radial coordinates --  the gradient of the red-shift function. As 
$\delta'(r)=\frac{-4\pi(\rho+p)}{\mu}$. As $\mu >0$ outside the horizon, so the sign of $\delta'(r)$ is 
governed by the sum of the energy density($\rho$) and radial pressure($p$).  
The negative values of $\delta$ in the whole ranges of radial coordinates confirm that 
the external matter satisfies the Weak Energy Condition (WEC), specifically $\rho+p \gsim 0$. 
This plot quantifies how the local matter density and pressure actively curve the space-time, 
modifying the time-dilation factor $g_{tt}$ away from the pure vacuum Schwarzschild solution. 
At the event horizon the regularity condition tells us that $\rho=0$, so thus $p$, therefore 
$\delta'$ is zero too. This leads to the constant values of $\delta(r_H)$ for three 
different values of the core-radius.

In the top-right panel of Fig.\ref{delta} we plotted the red-shift parameter (in logarithmic scale) 
as a function of the radial parameters. This effectively shows the order of magnitude of the metric 
correction. The values of the $\delta$ is approximately $10^{-1}$ to $10^{-3}$, this means that the 
matter-induced corrections are small but non-negligible 10.0\% to 0.1\% level respectively. For 
thicker halos ($r_0=20$) the correction grows steadily, indicating accumulated gravitational 
influence of the extended matter distribution. However for smaller core radius with $r_0=2.0$, 
due to the large mass accumulation near the event-horizon disk the red-shift parameter has 
appreciable effect with radial co-ordinates. Even though the values of the $\delta$ is small, 
it significantly affects ISCO location through the effective potential and the characteristic equations, 
Eqns.\ref{veff},\ref{vpeff},\ref{vppeff}.


In the bottom-left panel of Fig.\ref{delta} we plotted the double derivative of the 
red-shift function ($\delta''$). The $\delta' <0$ for $r_0$= $10.0$ and $20.0$ indicates 
$\rho+p >0$ (thus satisfies WEC). The matter density and pressure are positive and significant. 
The increasing trend shows matter effects dilute with increasing values of radial co-ordinate.  
These three parameters, i.e., $\delta,\delta',\delta''$ appear in the double derivate of the 
effective potential, in Eqn.\ref{vppeff}, thus much important for finding the $r_{ISCO}$.

In the bottom-right panel of Fig.\ref{delta} we have plotted the $g_{tt}=e^{-2\delta}\mu$ of the 
metric. This function governs the gravitational red-shift and the effective potential profile 
for stationary observers. Like Schwarzschild (bald) black-hole the values of $g_{tt}=0$ at the event 
horizon $r_H=2$ (due to regularity condition) for all the three values of the $r_0$. The values 
of $g_{tt}$ increases monotonically as $r$ increases and as $r\to\infty$ the values 
of $\delta \to 0$ and $\mu \to 1$, so $g_{tt}=e^{-2\delta}\mu$
leads to one. This ensures the asymptotic flatness of the metric far away from the black-hole. 

The parameter $r_0$ (i.e., the core-radius or the scale parameter in the density profile) characterizes 
the spatial extent in which the external matter field distributed. The plot demonstrates that the 
large extended matter distribution with $r_0 = 20.0$ yields a significantly lower value 
of $g_{tt}=e^{-2\delta}\mu$ at a given radius compared to a more compact distribution with $r_0 = 2.0$.
This suppression of $g_{tt}$ for larger values of $r_0$ indicates that a thicker ADM-halo  
modifies the space-time geometry more profoundly in the near-horizon region. 

\begin{figure}[ht!]
\begin{center}
\raisebox{0.0cm}{\hbox{\includegraphics[angle=0,scale=0.57]{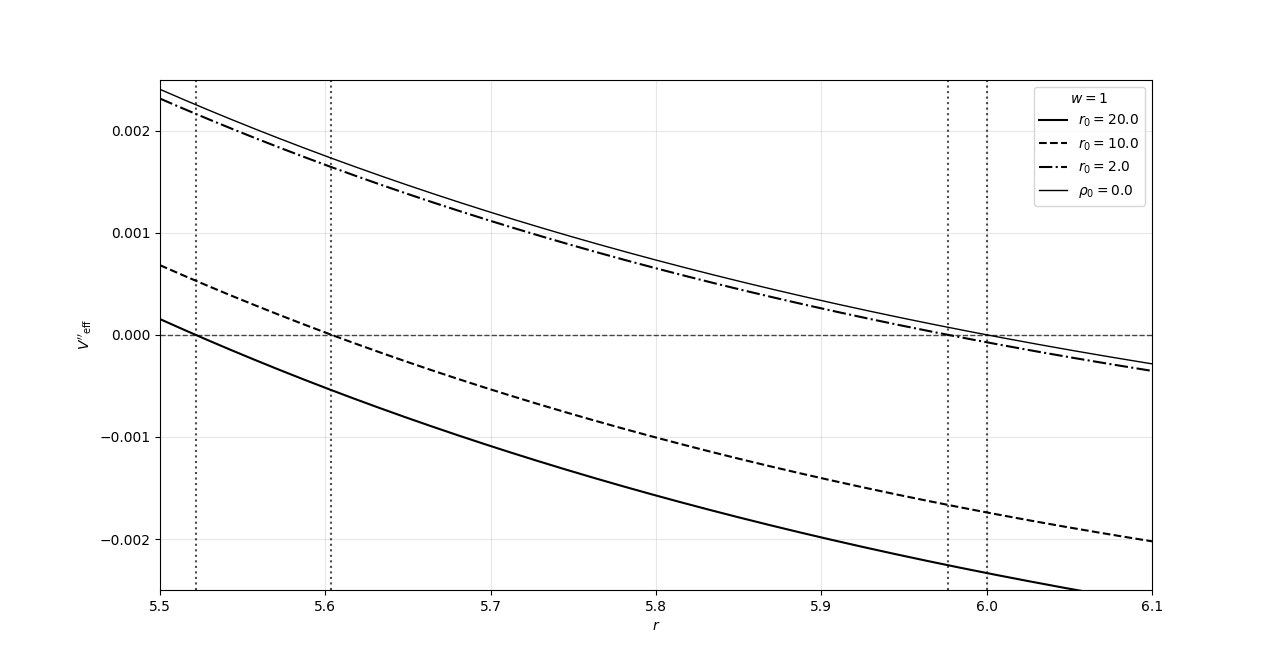}}}
	\caption{The profile of $V_{eff}^{''}(r)$ of the massive particle as 
	a function of radial coordinate~($r$) for three different values of 
	core radius $r_0$=$2.0,10.0$ and $20.0$ using dashed, 
	dotted-dashed and thick-solid respectively. The values of $r_{ISCO}$ are 
	those for which $V_{eff}^{''}(r)=0$. The Schwarzschild (bald) case with $\rho_0=0.0$ 
	is also plotted using the thin-solid line style which gives the maximum allowed upper limits 
	following~\cite{Cen:2025atm}.}
\label{photoniscow1}
\end{center}
\end{figure}

Therefore the larger values of $r_0$ effectively deepens the gravitational potential well 
relative to the vacuum Schwarzschild scenario. This is the reason for larger inward 
shift of the $r_{ISCO}$ for larger values of $r_0$ is observed in Fig.\ref{photoniscow1}.

We have plotted the profile of $V''_{\text{eff}}$ as a function of radial coordinate  
in Fig.\ref{photoniscow1} with three different values of the halo core radius 
$r_0=20.0,10.0.,2.0$ (using thick, dot-dashed and dashed line-style respectively). 
The standard bald Schwarzschild case (is where the ADM-halo energy density is equal to zero) 
is the one with $\rho_0=0.0$ plotted with thin solid line. 

It is important to note that the location of the stability of pure vacuum case ($\rho_0 = 0$) 
is significantly deviates from the ADM-halo scenario with three non-zero core-radius. This  
highlights how the external matter field alters the effective potential landscape near 
the horizon and creates a complex stability structures distinct from the pure vacuum solution. 

In Fig.\ref{photoniscow1} we have shown the exact location of the $r_{ISCO}$ 
of the matter-filled space-times, $r_0 = 2.0, 10.0, 20.0$, by putting the stability condition 
$V''_{\text{eff}}(r) = 0$. 
The thin solid line ($\rho_0 = 0$) represents the Schwarzschild vacuum, crossing $V''_{\text{eff}}(r) = 0$ 
exactly at $r = 6$. Our numerical adoption validates the analytical findings of 
theoretical upper limits of ($r_{\text{ISCO}} \le 6$) ~\cite{Cen:2025atm}. Our 
results demonstrate that the presence of the external matter field shifts the $r_{ISCO}$ inward. 
The magnitude of the shift depends upon the ADM-halo core radius, $r_0$. We found that 
for $r_0=2.0,10.0,20.0$ the numerical values of $r_{\text{ISCO}}$ are approximately 
$5.97$, $5.61$, $5.53$ respectively.

\section{Summary and Outlook}
\label{sec:summary}

In this analysis we have investigated the geodesic structure of a static, spherically symmetric 
black-hole in the background of a Astrophysical Dark Matter (ADM)-halo. The considered halo density 
profile is purely a phenomenological origin. The presence of $(1-r_H/r)$ factor 
guarantees the regularity at the event-horizon and the exponential factor ensures 
the finite total halo mass. We numerically integrated the Einstein field equations 
with equations of state $p=w\rho$ for the stiff matter $w=1$ case. We analyzed the 
modifications to the effective potential for time-like massive particle.

The central finding of our analysis is the rigorous numerical consistency of the recently proposed 
theoretical upper limit \cite{Cen:2025atm} for the ISCO radius, $r_{\text{ISCO}} \le 6$ with unit 
normalized black-hole mass. We showed that the presence of ADM-halo pulls the ISCO inward 
with respect to the vacuum Schwarzschild scenario with $\rho_0=0$. The magnitude of these inward 
shifts are highly sensitive to the spatial extent of the halo characterized by it's core radius 
$r_0$. A more diffused halo (i.e., larger $r_0$) induces a more pronounced deviation 
from the vacuum Schwarzschild scenario (i.e., bald black-hole geometry).

We would like to mention here that our analysis has direct and testable implications 
in high-energy observational astrophysics. The inward shift of $r_{ISCO}$ alters the inner  
edge of the accretion disk, potentially modifying the thermal X-ray emission spectra 
and the radiative efficiency of the black-hole. Furthermore, the modified geodesic structure 
will affect the gravitational waveforms emitted by compact objects in-spiraling into 
such hairy black-holes, offering a potential avenue for detection with LIGO/Virgo/KAGRA 
and LISA interferometry experiments.

In this present analysis the energy and angular momentum of the massive particles are held fixed. 
However it would be interesting to see the effects of varying the energy and angular 
momentum and also the case of radiation-like fluid with $w=\frac{1}{3}$ are 
left for future work.

\subsubsection*{Acknowledgments}  

ASK express sincere gratitude to the Great Lunar Expedition Mission, under the Colorado Space Grant 
Consortium (USA), for their invaluable institutional support and the inspiration they provided 
throughout this research. SPD acknowledges the High Performance Computing facility 
at Shivaji University Kolhapur.

\end{document}